# Realization of a sonic black hole analogue in a Bose-Einstein condensate

Oren Lahav, Amir Itah, Alex Blumkin, Carmit Gordon, Shahar Rinott, Alona Zayats, and Jeff Steinhauer

*Technion – Israel Institute of Technology, Haifa, Israel*

We have created an analogue of a black hole in a Bose-Einstein condensate. In this sonic black hole, sound waves, rather than light waves, cannot escape the event horizon. A step-like potential accelerates the flow of the condensate to velocities which cross and exceed the speed of sound by an order of magnitude. The Landau critical velocity is therefore surpassed. The point where the flow velocity equals the speed of sound is the sonic event horizon. The effective gravity is determined from the profiles of the velocity and speed of sound. A simulation finds negative energy excitations, by means of Bragg spectroscopy.

The event horizon is a boundary around the black hole, enclosing the region from which even light cannot escape. It has been suggested that an analogue of a black hole could be created in a variety of quantum mechanical [1-6] or classical [7-9] systems. In the case of a quantum fluid such as the Bose-Einstein condensate studied here [3], it is sound waves, rather than light waves, which cannot escape. This sonic black hole contains regions of subsonic flow, as well as regions of supersonic flow. Since a phonon cannot propagate against the supersonic flow, the boundary between the subsonic and supersonic regions marks the event horizon of the sonic black hole. The analogy was later extended to include excitations with a non-linear dispersion relation, in addition to phonons [10-12].

The experimental challenge is to create a steady flow which exceeds the speed of sound [1,3,13,14]. Consider a phonon with momentum $\hbar \mathbf{k}$. In the reference frame of the moving fluid, the phonon has energy $E = \hbar k c$, where $c$ is the speed of sound. In the laboratory frame, by a Galilean transformation [15], this energy becomes $E' = E + \hbar \mathbf{k} \cdot \mathbf{v}$, where $\mathbf{v}$ is the flow velocity. For the case of supersonic flow ($v > c$), $E'$ can be zero, resulting in the unstable production of phonons. This instability is thought to prevent the supersonic flow required to realize a sonic black hole, a phenomenon referred to as the Landau critical velocity [3,13,15]. By momentum conservation however, the production of such phonons requires an additional body such as an impurity particle [16] or a container with a rough wall [15]. This body provides momentum in the opposite direction to the flow. Thus, we have arranged an experimental apparatus which does not supply much momentum in this direction, allowing for supersonic flow during the timescale of the experiment [3]. The free flow required to overcome the Landau critical velocity also helps prevent the production of quantized vortices, which usually limit the flow to speeds much lower than the speed of sound [17].

Suggested schemes for forming a sonic black hole in a condensate include a Laval nozzle [18, 19], flow along a ring or a long, thin condensate [3, 20], a gradient in the coupling constant [21,22], a soliton [2,23], an expanding condensate [24], and repulsive potential maxima [5,25]. We achieve the black hole horizon by a step-like potential combined with a harmonic potential, as shown in Fig. 1. We translate the harmonic potential to the left as indicated by the horizontal arrow, moving the condensate towards the stationary step. While crossing the step, the condensate accelerates to supersonic speeds. Thus, the region to the left of the step is supersonic, and the region to the right is subsonic. There is therefore a black hole horizon at the location of the step.

The condensate consists of $1 \times 10^5$ $^{87}$Rb atoms in the $F = 2$, $m_F = 2$ state, and is initially prepared in the harmonic part of the potential, a magnetic trap with oscillation frequencies of 26 Hz and 10 Hz in the radial and axial ($y$) directions respectively. The $x$-coordinate of the minimum of the harmonic trap is controlled by adjusting the trap frequencies, which adjusts the sag due to gravity (in the -$x$ direction). The step-like potential is created by a large diameter, red-detuned laser beam with a Gaussian profile (1/e$^2$ radius of 56 µm, wavelength 812 nm). Half of this beam is blocked, so that the boundary between the dark and light regions forms the potential step of height $V_0 / h = 2.0 \, \text{kHz}$. Initally, the condensate is located to the right of the step, as shown in Fig. 1. Starting at $t = 0$, the harmonic potential is accelerated until it reaches the constant velocity of roughly 0.3 mm s$^{-1}$. The condensate then passes over the potential step, as shown in Figs. 2a and 2b. We observe no increase in the thermal fraction in this process. Furthermore, it is seen that the density of the condensate is much smaller to the left of the potential step, whose location approximately coincides with the dashed lines in Figs. 2a and 2b. By conservation of mass, the decrease in density corresponds to an increase in flow velocity.

The images of Fig. 2a and 2b can be converted into density profiles, averaged over the cross section of the condensate, as shown in Fig. 2f. This average density is found by the relation $n = N / \delta x A$, where $N$ is the number of atoms in a column of pixels of Fig. 2a and 2b, $\delta x = 0.46 \,\mu\text{m}$ is the length of a pixel, and $A$ is the cross-sectional area of the condensate in the *y-z* plane, seen in Figs. 2c and 2d. To calibrate $N$, the sensitivity of the imaging system is required. This sensitivity is found by studying the condensate in the harmonic potential only, shown in Fig. 2e. The observed radius and length of this condensate, combined with the known trap frequencies, give the total number of atoms $N_C$ in the condensate [26]. $N_C$, divided by the sum of the pixels of Fig. 2e, gives the sensitivity of the imaging system.

By comparing the two curves of Fig. 2f, it is seen that the density in the left region of the figure increases with time, while the density in the right region decreases. We can use this redistribution of density to compute the flow velocity via the continuity equation, given by $\nabla \cdot (n\vec{v}) = -\partial n / \partial t$ [26]. In Figs. 2a and 2b, the flow appears to be largely in the *x*-direction. This unidirectional flow is further verified by Figs. 2c and 2d, which show no dynamics in the *y-z* plane. The shape in this plane remains similar to the shape of the condensate in the harmonic trap, shown in Fig. 2e. Assuming that the flow is in the *x*-direction only, the continuity equation gives

$$v = -\frac{1}{n} \int_0^x \frac{\partial n}{\partial t} dx'. \tag{1}$$

The integrand of Eq. 1 is given by the ratio $\Delta n / \Delta t$, where $\Delta n$ is the difference between the profiles in Fig. 2f (after normalizing them to the same total atom number), and $\Delta t$ is the time difference between the profiles. This integrand is discretized by the pixels of the imaging system. The origin $x = 0$ is the left edge of Fig. 2f. Fig. 2g shows the resulting velocity profile. This calculation is repeated for various times, giving the velocity profiles of Fig. 3a. The error bar indicates the standard error of the mean. We attribute this variability to fluctuations in the initial position of the harmonic magnetic trap. The experimental results agree well with a 3D simulation of the Gross-Pitaevskii equation shown in Fig. 4b.

For comparison with $v$, the profile of the speed of sound can be computed from the density profiles of Figs. 2f and 2h, via the relation $c = \sqrt{g n_{ave} / m}$ [27], where $g$ is the interaction parameter, $m$ is the atomic mass, and $n_{ave}$ is the average of the several density profiles shown in the figure. In contrast to $v$, the computation of $c$ relies on the absolute calibration of $n$ discussed above. The profiles of $c$ are indicated by the solid black curves in Figs. 2g and 3a.

As seen in Fig. 3a, the peak flow velocity $-v_m$ exceeds $c$ by an order of magnitude. The black hole horizon $x_H$ is indicated by a filled circle. The flow decreases below the speed of sound again at the white hole horizon, indicated by a "+" [3]. In order to find the position-dependent profiles of the horizons, $v$ and $c$ are computed for 3 horizontal slices in Figs. 2a and 2b, rather than for the entire image. The dashed and dash-dotted lines indicate the resulting profiles of the black hole horizon and white hole horizon, respectively.

The black hole analogy requires that $v$ and $c$ be stationary (independent of time). We find that the motion of the black hole horizon itself is a reasonable overall check of stationarity. Specifically, the speed $v_H$ of the black hole horizon should be much less than $c$. In Fig. 3a, the total change in $x_H$ over 20 ms gives $v_H = 0.12c$. Thus, the flow is almost stationary, even in the laboratory frame.

In analogy with a black hole which traps photons, the supersonic region of a sonic black hole can trap a range of Bogoliubov excitations [3]. Due to the flow, the group velocity measured in the laboratory frame is decreased by $-v$. An excitation is trapped (dragged away from the horizon) when it has a negative group velocity. Due to the non-linear dispersion relation, excitations with very short wavelengths can certainly escape. The minimum trapped wavelength is approximately $-2\pi\hbar/mv_m = 1.6\,\mu m$. However, the wavelength of the excitation must be shorter than the 18 µm width of the supersonic region between the horizons [3]. Thus, wavelengths between 1.6 µm and 18 µm are trapped. Phonons are among the trapped excitations, since phonons have wavelengths greater than $2\pi\xi = 5.2$ µm, where $\xi$ is the healing length.

Finally, we address the possibility of using the sonic black hole for a future study of analogue Hawking radiation [28]. Hawking radiation requires that the trapped excitations have negative energy, which we verify by simulating Bragg spectroscopy. In the simulation, a pair of far-detuned laser beams are focused onto the supersonic region, as indicated in the inset of Fig. 4d. There is a slight frequency difference between the beams, given by $\omega = \omega_1 - \omega_2$. If the condensate absorbs a photon from beam 1 and emits a photon into beam 2, then an excitation with energy $\hbar\omega$ is created. The momentum $\hbar k$ of the excitation is in the *x*-direction, and is predetermined by the angle between the beams. The solid curve of Fig. 4d shows the momentum $\Delta p$ transferred to the condensate during a 2 ms pulse of the Bragg beams. The peak *B* corresponds to absorbing photons from beam 1, since $\Delta p > 0$. However, $\omega < 0$, so the excitation indeed has negative energy. Fig. 4e shows the dispersion relation, the locations of the peaks as a function of *k*. The green curve indicates the trapped excitations with negative energy.

Further insight into the negative energy excitations can be obtained by looking at the wavefunction, Fig. 4f. The high frequency oscillations in the supersonic region correspond to a well-defined momentum. This momentum Doppler-shifts peak *B* of Fig. 4d to negative energies, in contrast to peak *C* of the subsonic region. It is interesting to note that a grey soliton has *v* and *c* profiles which are qualitatively similar to Fig. 3a. However, the wavefunction has no oscillations [29], so Bragg spectroscopy would find no negative energy excitations for a grey soliton.

Hawking radiation was first considered for excitations in the phonon regime [1]. This computation relied on a negligible quantum pressure term in the flow equation [1,3,26,30]. The simulation shows that this term is indeed much smaller than the other terms, as seen in Fig. 4c.

The effective temperature of Hawking radiation is given by $k_B T_H = \hbar g_H / 2\pi c_H$ [1], where $c_H$ is the speed of sound at the black hole horizon, and the effective surface gravity is given by $g_H = c_H(dc/dx + dv/dx)|_{x=x_H}$ [18]. The opposite sign of the slopes seen for $-v$ and $c$ in Fig. 3a serves to increase $g_H$ and $T_H$, which are shown in Fig. 3b. In order for the Hawking radiation to have a thermal spectrum, the flow should be sufficiently hydrodynamic, meaning that the gradients in $v$ and $c$ should be sufficiently small [11,22]. The relevant parameter is $\lambda$, which is the number of healing lengths required for $c+v$ to change by $\sqrt{2}c$. For the present work $\lambda = 7$, which is sufficiently large to give an approximately thermal spectrum with an effective temperature close to $T_H$.

We can write the Hawking temperature as

$$k_B T_H = \frac{\mu}{\pi \lambda} \qquad (2)$$

Where $\mu = mc^2$ is the chemical potential of the condensate. This implies that $k_B T_H$ should be somewhat less than $\mu$. A thermal cloud with such a low temperature might be difficult to differentiate from the condensate or from other excitations [31]. Therefore, it might be useful to employ correlation techniques [11,21,22], or perhaps the region of Fig. 3a between the white and black hole horizons could serve as a resonator for the lasing of phonons, which could amplify the Hawking radiation [32]. The Hawking radiation should have a wavelength shorter than the width of the supersonic and subsonic regions [3]. This implies $T_H > 2$ nK. In order to observe Hawking radiation, the width should be increased, or $T_H$ should be increased by an order of magnitude (see Fig. 3b). By Eq. 2, the latter can be achieved by increasing the density, as well as the gradients in $v$ and $c$. Furthermore, it is seen in Fig. 3b that the black hole horizon is maintained for at least 20 ms. This time period corresponds to at least one period of 2 nK radiation. Thus, the lifetime of our sonic black hole should be sufficient or almost sufficient for observing Hawking radiation. In addition, Fig. 3a shows that the white hole horizon is stable for the duration of the experiment [3,14].

In conclusion, a sonic black hole/white hole pair has been realized in a Bose-Einstein condensate. The supersonic flow field results from a step-like potential. The Landau critical velocity is exceeded by an order of magnitude. The effective surface gravity is extracted from the profiles of the in-situ velocity and the speed of sound. The flow field, combined with increased velocity gradients and density, could potentially be used in a study of Hawking radiation.

We thank Amos Ori, Yariv Kafri, Rami Pugatch, James Anglin, Renaud Parentani, Jean Macher, Serena Fagnocchi, Iacopo Carusotto, Andrea Trombettoni, Ehud Altman, and Emil Polturak for helpful discussions. This work is supported by the Israel Science Foundation.

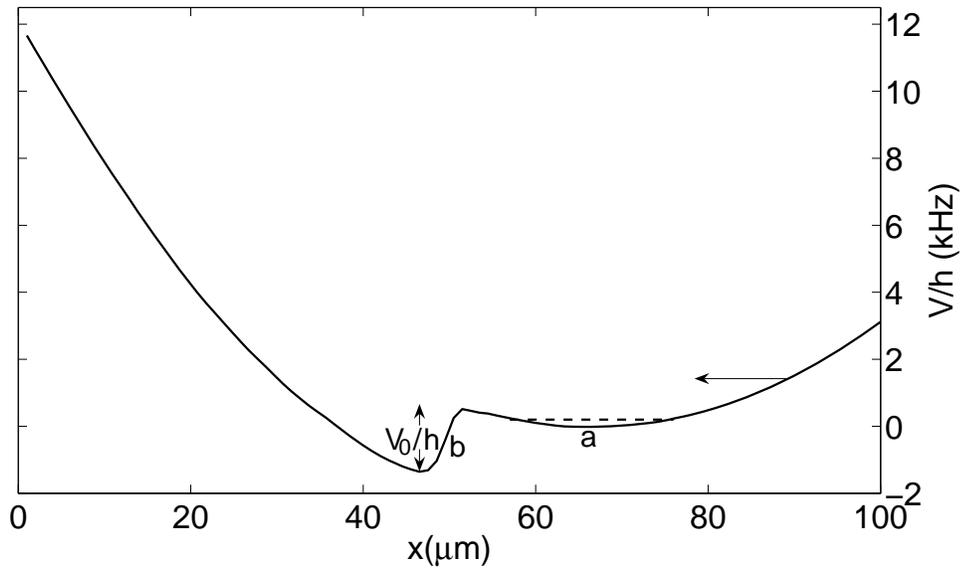

FIG. 1. The profile of the potential, shown at a time before the harmonic potential (a) has reached the step-like potential (b). The horizontal arrow indicates the motion of the harmonic potential relative to the stationary step-like potential. The dashed line indicates the chemical potential of the condensate. Gravity is in the -$x$ direction. The profile of the potential step is derived from an image of the laser beam. The harmonic potential is derived from the measured frequency of the harmonic magnetic trap.

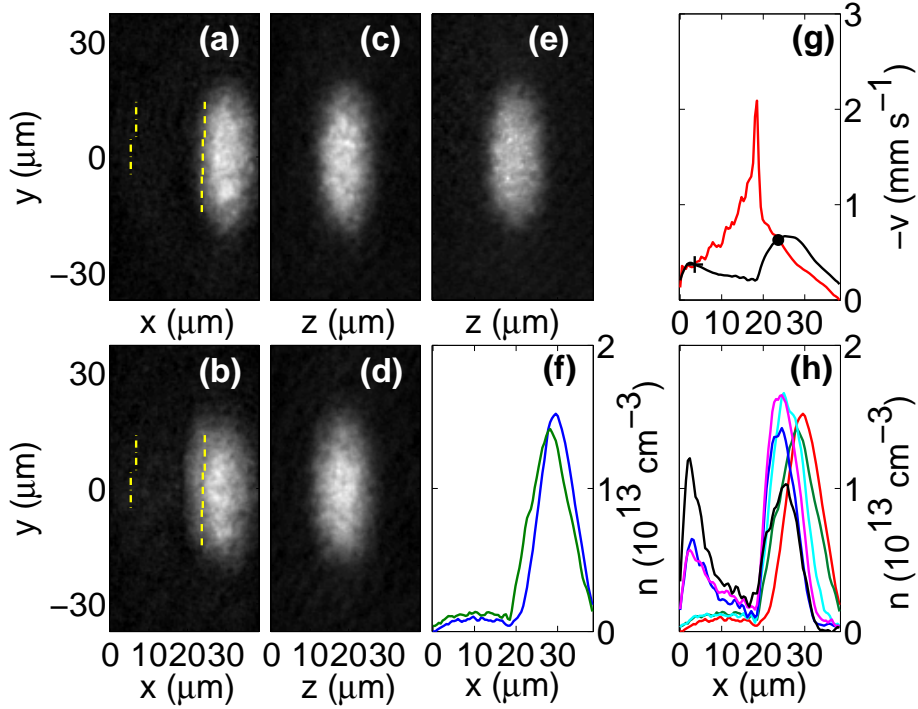

FIG. 2. In-situ absorption images of the sonic black hole, with the potentials on. Gravity is in the -x direction. (a) Condensate in the presence of the step-like potential, at 200 ms. The average of 2 images is shown. The dashed (dash-dotted) line indicates the position-dependent profile of the black hole (white hole) horizon. (b) Like (a), at 205 ms. (c) and (d) Side views (y-z plane) of (a) and (b), imaged simultaneously with (a) and (b). (e) Side view of the condensate in the harmonic potential only. (f) Density profiles, derived from (a) and (b) in blue and green, respectively. The density is averaged over the y-z plane (see text). (g) -v (red curve) and c (black curve). These curves are derived from (f), as described in the text. The filled circle and plus indicate the black hole and white hole horizons, respectively. (h) Time evolution of the density profile. The red, green,

cyan, blue, magenta, and black curves indicate equal intervals from 200 ms to 225 ms. Each curve is derived from an average of 2 images.

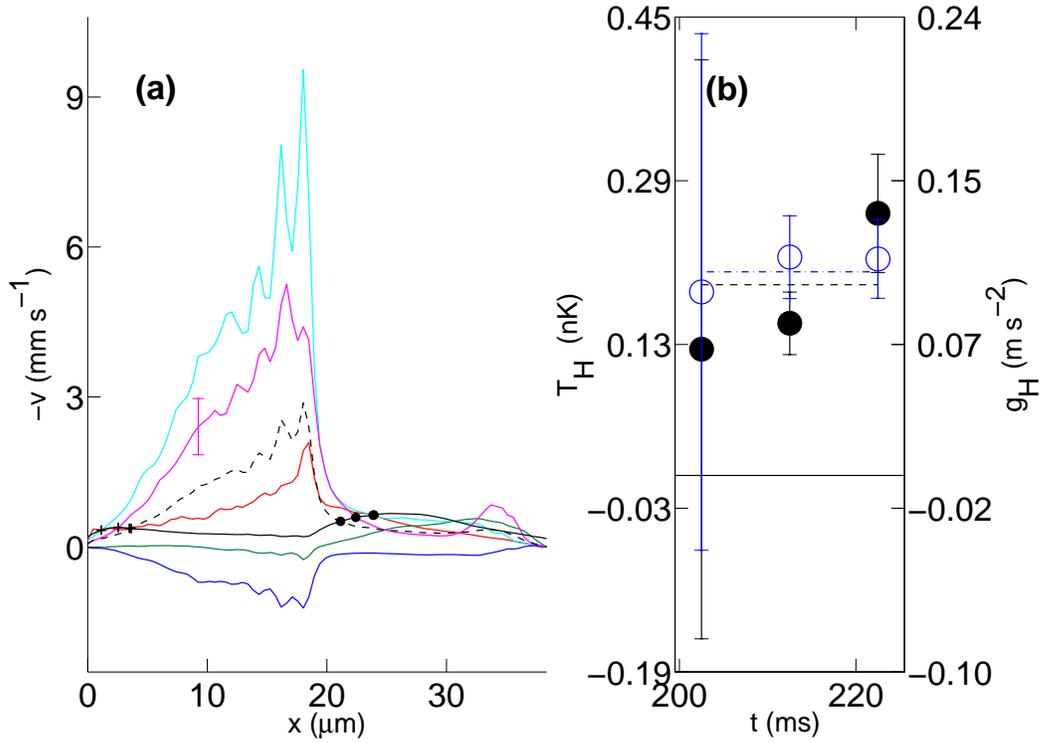

FIG. 3. Time evolution of the flow velocity. (a) The flow velocity versus the speed of sound. The red, green, cyan, blue, and magenta curves show $-v$ at equal intervals from 202.5 ms to 222.5 ms, derived from adjacent curves in Fig 2h (see the text). The dashed curve indicates the average $-v$ from 202.5 ms to 222.5 ms. The black solid curve indicates the average $c$ (see the text). The filled circles (pluses) indicate the black hole (white hole) horizon. The fringes near the velocity maximum are an artifact of the imaging. (b) Hawking temperature and effective surface gravity. The filled and open circles correspond to the entire condensate and the central slice, respectively. The dashed curve corresponds to the average (dashed) velocity curve of (a). The dash-dotted curve corresponds to the average velocity curve of the central slice of the condensate.

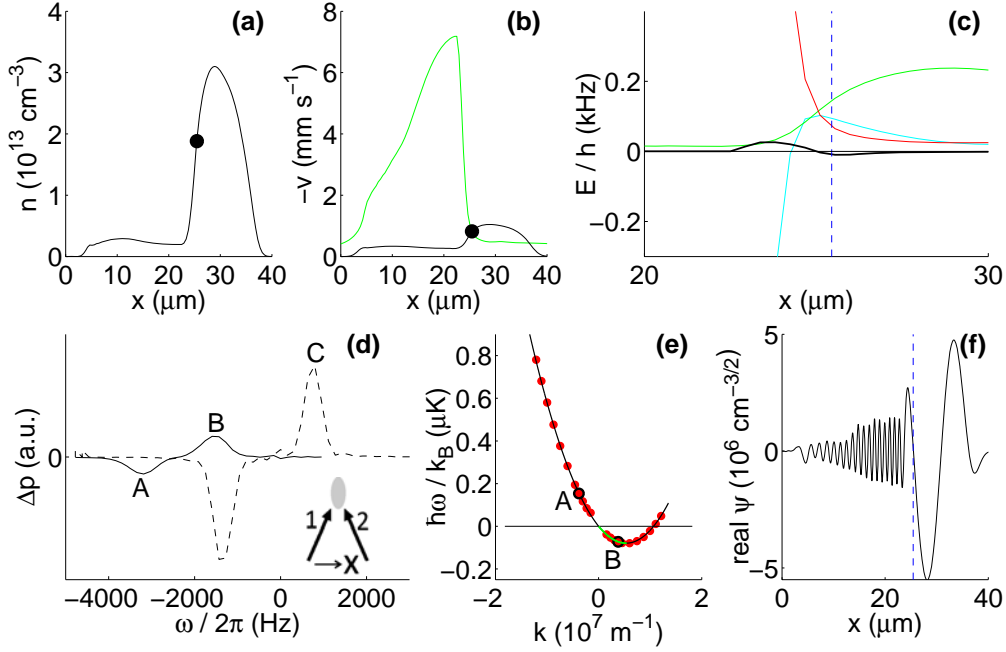

FIG. 4. 3D Gross-Pitaevskii simulation of the experiment. The central axis of the condensate is shown. (a) The density. The filled circle indicates the black hole horizon. (b) The flow velocity versus the speed of sound. The green curve shows $-v$ and the black curve indicates $c$. Here, the simulation is analyzed with the same technique and resolution as in Fig. 3a from the experiment. (c) The energy terms in the flow equation. The thick black curve shows the quantum pressure term $-\hbar^2/2m(\nabla^2\sqrt{n}/\sqrt{n})$. The green curve shows the interaction term $gn$. The red curve shows the inertial term $mv^2/2$. The cyan curve shows the potential term $V$. The latter two terms are off-scale for much of the figure. The dashed line indicates the black hole horizon. (d) The momentum transferred

by a Bragg pulse. The solid (dashed) curve corresponds to the supersonic (subsonic) region. Peaks A and B correspond to A and B in (e). The inset shows the two Bragg beams incident on the condensate. (e) The dispersion relation in the supersonic region. The circles result from Bragg spectroscopy. The solid line is the Bogoliubov dispersion relation. The green curve indicates the trapped excitations with negative energy. (f) The real part of the wavefunction.